\documentclass{article}
\usepackage{epsfig}
\usepackage[centertags]{amsmath}
\usepackage{graphicx}
\begin{document}

\title{Chaos assisted adiabatic passage }

\author{Kyungsun Na and L. E. Reichl
\\ Center for Studies in Statistical Mechanics and Complex Systems,
\\ The University of Texas at Austin, Austin, TX 78712 USA }

\maketitle
\begin{abstract}

We study the exact dynamics underlying stimulated Raman adiabatic
passage (STIRAP) for a particle in
a  multi-level anharmonic system (the infinite square-well) driven by
two sequential laser pulses,
each  with constant carrier frequency.  In phase space regions where
the laser pulses
create chaos, the particle can be transferred coherently into energy states
different from  those predicted by traditional STIRAP. It appears
that a transition to chaos can
provide a new tool to control the outcome of STIRAP.

\end{abstract}

\section{Introduction}

Laser radiation provides a means to control
intra-molecular processes in a robust manner because of a conservation law
that comes into play when monochromatic radiation interacts with
nonlinear systems.
The origin of this conservation law  is the
discrete time-translation invariance of laser driven systems.
For radiation interacting with molecular systems, this conservation
law  gives rise to stable electron-photon (phonon-photon,
roton-photon) structures described by conserved eigenstates (the
Floquet states) of the
driven system. Floquet states are exact eigenstates of periodically driven
systems \cite{shirley}, \cite{sambe}, \cite{reichl}. Their reality
can be seen in recent
atom-optic experiments \cite{steck}, \cite{hensinger}, \cite{luter},
\cite{reichl}, where
millions of sodium \cite{hensinger} or cesium atoms \cite{steck}, interacting
with a time-modulated standing wave of light, underwent large coherent periodic
oscillations in momentum in a multi-photon process. These coherent
oscillations were
subsequently found to be due to the interference of only two or three
Floquet eigenstates
whose phase space structure was determined by an underlying chaotic sea
induced by the interaction between the atoms and the light
\cite{luter}. In this paper, we
wish to show that similar mechanisms are important when laser pulses
interact with
intra-molecular forces. We will focus on the exact dynamics
underlying STIRAP (stimulated
Raman adiabatic passage) for laser pulses interacting with a simple model of
intra-molecular dynamics.

STIRAP  has become an important tool for coherently controlling and
changing the
vibration and electronic states of entire molecular populations with
close to 100$\%$
efficiency.  STIRAP involves the
application of short laser pulses with carefully chosen carrier
frequencies to a
molecular system for the purpose of exciting the molecules in a controlled
manner. This technique causes a coherent change in the entire
molecular population
between molecular states.  Traditional models of
STIRAP  generally view the molecular target  as a simplified
multilevel system. Indeed, traditional STIRAP focuses on three carefully chosen
vibration and/or electronic levels. There are no studies, that we
know of, that look
at the effect on STIRAP of the actual nonlinear dynamics that occurs
when the laser pulses
interact with a molecular system. However, we know that laser pulses
can induce chaos
and this can strongly affect the response of the molecule to the laser field.
In addition the internal dynamics of the molecule may itself be exhibiting
the manifestations of chaos and simple pictures of the molecular level
structure are likely not valid. Understanding this dynamics is very
important for extending STIRAP  to wider ranges of molecules

In this paper we study the {\it exact dynamics} underlying STIRAP for
a model system that
contains the essential features of low energy vibration states, or
rotational states, of
a diatomic molecule driven by two laser pulses.  We will find, for
example, that in phase
space regions where the laser pulses create chaos, the molecule can
be transferred
coherently into energy states very different from those predicted by
traditional STIRAP.
It appears that a transition to chaos may provide a new tool to
control the outcome of
these processes in molecular systems.

STIRAP was first proposed by Hioe \cite{hioe}, \cite{oreg} and later
confirmed in an
experiment involving population transfer between vibration-rotation
states of sodium
dimers \cite{gaubatz1}, \cite{gaubatz2}.  There are several
variations to STIRAP which
generally is described as a process involving three particular energy levels,
$E_1,E_2$, and $E_3$,  of a collection of atoms or molecules of interest.
All atoms or molecules
are initially in the lowest energy state $E_1$. Two laser pulses then
impinge sequentially on the system in order to make a transition toward
the target state $E_3$.
In the ``ladder" version of STIRAP the target state $E_3$ is the highest state
in energy and  in the
``lambda" version of the STIRAP  the intermediate state $E_2$ is the highest
state in energy.
The first laser pulse couples $E_2$ and $E_3$  which contain no population, and
the second laser pulse then couples $E_1$ and $E_2$.
As a result of these processes, the entire
  atomic or molecular  population  is transferred coherently into the
target state without populating the intermediate state
after the laser pulses have passed
\cite{shoreA},
\cite{shoreB}, \cite{dani},
\cite{berg}, \cite{yat}, \cite{drese}, \cite{ye}.

As mentioned above, conventional STIRAP analyzes
an atomic or  molecular system in terms of  three carefully chosen
energy levels of the
unperturbed system.  However, complications arise in real applications
since these systems are composed of multiple states due to the
rotational, vibration,
and electronic levels of the unperturbed system.
Extension of STIRAP to multilevel systems has been discussed by a
number of authors,
generally in terms of simplified models using the rotating wave
approximation and three or more laser pulses
\cite{shore,tannor,hioe1,oreg1,smith,band,hioe2,martin1,martin2,broers,chelkowski,chang}.
   Alternatively, Raman chirped adiabatic passage schemes
\cite{broers,chelkowski,chang}, in which the frequency of the
incident laser pulse is
continuously changed, have also been proposed to allow the system to
climb through a
sequence of molecular energy levels.

   In this paper we take a different approach to the problem. We
consider the {\it exact
dynamics} that takes place when the two laser pulses impinge on a
multi-level system.
Instead of isolating a certain number of  levels of interest and
analyzing the process in terms of the rotating
wave approximation,
   we will let the
full multilevel dynamics evolve and  allow the system to tell us
how many levels we
must keep to accurately describe the atomic or molecular dynamics
when a radiation field
is present.  Unlike the previous studies which utilize more than two
pulses in multi-level
systems, we only apply two pulses and  allow as many
levels as dictated by the dynamics participate in the process.  One 
way to see how
many levels must be
kept is to look at the underlying classical phase space of the system
of interest. The
laser field, when interacting with nonlinear intra-molecular dynamics
will induce
nonlinear resonances and chaos in localized regions of the phase
space. Those structures
in the classical phase space which have a size greater than Planck's
constant determine
the structure of the Floquet eigenstates of the system, and thereby
have a direct
influence on the STIRAP process.

The model we use to study the effect of chaos on STIRAP in a
multi-level system is that of
a particle in an infinite square-well potential \cite{lin}, \cite{gald}.  The
infinite square-well potential is an anharmonic potential of the form
$x^{2n}$ in the
limit $n\rightarrow\infty$. It can give some insight into the
behavior of low lying
vibration states, or rotation states, of molecular systems in the
presence of
sequential laser pulses. An approximate version of the square-well
potential we consider
here could also be constructed in an atom-optics experiment \cite{luter} or in
semi-conductor heterostructures. The dynamics of a particle in an
infinite square-well
potential is also interesting because the laser pulses can cause the
low energy particle
states to undergo a transition to chaos  \cite{chism}, \cite{timber}.
For the case of
monochromatic laser fields, this transition to chaos is accompanied
by a plateau of high
harmonic radiation whose cut-off is determined by the width in energy
of the chaotic sea
induced by the laser field \cite{chism}.

In the sections below, we describe the behavior of a particle in an
infinite square-well potential which is driven by two sequential
laser pulses whose
carrier frequencies are monochromatic and chosen
to couple specific unperturbed energy levels of the particle in the
square-well potential. We will find that for the case when the pulse
amplitudes vary
slowly in time so that the adiabatic theorem \cite{messiah} is
satisfied, the dynamics can
be described in terms "snapshots" of the underlying classical phase
space at selected
times as the laser pulses pass through the system.  In Sect. 2, we
describe the classical
dynamics that results from the laser pulses. In Sect. 3, we discuss
how we will describe
the quantum dynamics for the driven system. In Sect. 4 we introduce
Floquet theory. In
Sects. 5, 6 and 7 we show that a dynamics quite different from that of the
traditional STIRAP ladder model can occur  in multilevel systems due to the
presence of chaos, even for fairly weak pulse amplitudes.  Finally in
Sect. 8 we make some
concluding remarks.

\section{Classical dynamics}

The model system for this study is a particle located in an infinite square
well potential with spatial width, $2a$. The potential energy has the form,
$V(\tilde x)=0$ for $|\tilde x|<a$
and $V(\tilde x)=\infty$
for $|\tilde x|\geq a$, where ${\tilde x}$ is the position of the particle.
The classical Hamiltonian which describes
the  dynamics of a particle of mass $m$ moving in the potential well and
driven
by two sequential pulses of monochromatic radiation is  given by
\begin{equation}
\tilde H=\frac {\tilde{p}^2}{2m} + {\tilde U_f(\tilde t)} {\tilde x}
{\rm cos}({\tilde \omega_f}{\tilde t}) + {\tilde U_s(\tilde t)}
{\tilde x} {\rm
cos}({\tilde
\omega_s} {\tilde t}),
         ~~~~~ {\rm for}~~~~ |\tilde x| < a,
\end{equation}
where  $\tilde p$ is the  momentum of
the particle,
$\tilde t$ is the time, and ${\tilde \omega}_{f}$ and ${\tilde
\omega}_{s}$ are the
carrier frequencies of the first and second pulses, respectively. The
amplitude of the first (second) pulse at time ${\tilde t}$ is
${\tilde U}_f({\tilde t})$  (${\tilde U}_s({\tilde t})$).    If we
rescale parameters
using $\tilde
x=xa$, $\tilde p=p\hbar/a$,
$\tilde U_o=U_o \hbar^2/2 m a^2$, $\tilde t= 2 m a^2
t/\hbar$,  ${\tilde H}=H{\hbar}^2/2ma^2$, and
$\tilde \omega_{f,s} = \omega_{f,s} \hbar/2 m a^2$, where $\hbar$ is 
Planck's constant, then the Hamiltonian is
\begin{equation}
H = p^2 + U_f(t) x {\rm cos}(\omega_f t)
               + U_s(t) x {\rm cos}(\omega_s t),  ~~~~ {\rm for}~~~~ |x| < 1
\end{equation}
and all parameters are dimensionless. The energy has been re-scaled
in units of $\hbar^2/2 m a$ in order to
make comparisons with the corresponding quantum system
in subsequent sections.

The amplitudes $U_{f}(t)$ and $U_{s}(t)$ have Gaussian shape of the form,
\begin{equation}
U_{f}(t)=U_{o} {\rm exp}(- \beta (t - t_{f})^2 )~~~ {\rm and}~~~
U_{s}(t)=U_{o} {\rm exp}(- \beta (t - t_{s})^2 ),
\end{equation}
where $t_f<t_s$.
We can control the duration of each pulse by adjusting the parameter $\beta$
and we can control the amount of overlap of the
two pulses by changing $t_{f}$ and $t_{s}$. For simplicity, we assume that the
maximum amplitude $U_{o}$ and the width $\beta$ of the two
pulses are the same.
A schematic picture of the variation in time of the amplitudes of the
two pulses is displayed in Figure 1.
The first pulse  is turned on and drives the particle in the square
well and then, with an
appropriate delay time, the  second pulse is turned on. The whole
pulse sequence ends at the total
pulse duration time $t=t_{tot}$. For the purpose of marking time
intervals in our subsequent
discussion, we choose times, $t_1={1\over 20} t_{tot}$, $t_c={1\over 
2}t_{tot}$, and
$t_2={19\over 20} t_{tot}$.

We will be interested in how the classical phase space behaves in the
{\it adiabatic limit}
where the pulses are turned
on and off very slowly relative to certain intrinsic time scales in
the system, such as the periods of
the carrier frequencies. In this limit the amplitudes, $U_f(t)$ and
$U_s(t)$ remain essentially
constant during time intervals where the cosine terms oscillate many
times. We can get an idea of the
structure of the phase space by plotting a Poincare  surface of
section of the phase space for fixed pulse
amplitudes \cite{reichl}.
Thus we also consider the Hamiltonian
\begin{equation}
H = p^2 + U_f(t_{fix}) x {\rm cos}(\omega_f t)
               + U_s(t_{fix}) x {\rm cos}(\omega_s t),  ~~~ {\rm for}~~~|x| < 1,
\label{tpham}
\end{equation}
where the amplitudes are set to constants by choosing their value at
some fixed time $t=t_{fix}$.
The Hamiltonian in Eq. \ref{tpham} is time-periodic and we can view
the behavior of
the phase space using Poincare surfaces of
section \cite{reichl}.  The Poincare surfaces of section for time-periodic
Hamiltonians are strobe plots
of $p$ and $x$, i.e. plots of $p$ and $x$  each time the Hamiltonian
goes through one complete oscillation in time.

We can perform a canonical transformation to action-angle variables
$(J,{\theta})$ defined
$J=2 |p|/\pi$, and  $\theta=\pm \pi (x+1)/2$.  The Hamiltonian then
has the form,
\begin{equation}
\begin{aligned}
H=\frac{\pi^2 J^2}{4}&-\frac {4 U_f(t_{fix})}{\pi^2}
\sum_{\nu=-\infty}^{\infty}
\frac{1}{(2\nu-1)^2} {\rm cos}((2\nu-1)\theta-\omega_f t) \\
                            &-\frac {4 U_s(t_{fix})}{\pi^2}
\sum_{\nu=-\infty}^{\infty}
\frac{1}{(2\nu-1)^2} {\rm cos}((2\nu-1)\theta-\omega_s t), ~~~{\rm for}~~~
0{\leq}\theta{\leq}\pi.
\end{aligned}
\label{classicalH}
\end{equation}
An infinite number of nonlinear resonances are produced in the classical phase
space by the external fields.
The {\it primary resonances} are located at $J=J_\nu \equiv 2
\omega_{f,s} /((2\nu-1)
\pi^2)$. As $\nu$ increases,
the energy at which higher order primary resonances are located
decreases.

In Figure 2, we show strobe
plots of the classical phase space for the case with $U_{o}=3.0$. The 
commensurability of these frequencies will allow us to use Floquet 
theory when we analyze the quantum system. We  choose the pulse 
carrier frequencies to be
$\omega_f=3{\pi}^2/4$ and $\omega_s=5{\pi}^2/4$.
For the frequencies we have chosen, the period of the Hamiltonian is
$T_o=8/{\pi}$. For the five cases
shown in Figs. 2.a-2.e,  we fix the amplitude of the pulses by setting
(a) $U_{f,s}(t_{fix}){\equiv}U_{f,s}(t_1)$, (b)
$U_{f,s}(t_{fix}){\equiv}U_{f,s}(t_f)$, (c)
$U_{f,s}(t_{fix}){\equiv}U_{f,s}(t_c)$,
(d) $U_{f,s}(t_{fix}){\equiv}U_{f,s}(t_s)$, and  (e)
$U_{f,s}(t_{fix}){\equiv}U_{f,s}(t_2)$, respectively.
       For each of these choices of amplitude
$U_{f}$ and $U_s$,   we show strobe plots of the
classical phase space allowing the time dependence of the cosine
waves to vary.   The three largest primary
resonances ($\nu=1, 2, 3$) induced by the first pulse   are located
at $J=2.5$, $J=0.83$ and $J=0.5$, respectively.  The three largest primary
resonances ($\nu=1,
2, 3$) induced by the second pulse are located at $J=1.5$, $J=0.5$ and
$J=0.3$, respectively.   In Fig. 2.a, with  $U_f(t_1)=0.1667$ and
$U_s(t_1)=0.000003$, the primary resonances induced by the first pulse
are dominant.
In Fig. 2.e, with  $U_f(t_2)=0.000003$ and $U_s(t_2)=0.1667$, the
primary resonances induced by
the second pulse are dominant.
In all cases, the first primary resonance ($\nu=1$) is
located at the highest energy and
the higher order primary resonances are located at decreasing energy
as $\nu$ increases. As a result,
this system will always have a chaotic region at low energy due to
the overlap of higher order resonances.
For energies above the region of influence of the $\nu=1$
primary of the first pulse, the phase space is dominated by KAM
(Kolmogorov-Arnold-Moser) tori.

In Fig. 2.c, where $t_{fix}=t_c={1\over  2}t_{tot}$,
the primary $\nu=1$ resonances due to the two pulses have equal
amplitude and are clearly visible at
$J=2.5$ and $J=1.5$. For this case the pulse amplitudes
are $U_f(t_c)=U_s(t_c)=1.103$.
All the higher order primary resonances have been destroyed and a
large chaotic sea has formed at low energy.

Fig. 2.b shows the classical phase space at time, $t_{fix}=t_f$ when the
first  pulse reaches its maximum amplitude with
$U_f(t_f)=3.0$ and $U_s(t_f)=0.055$.  The region of phase space about
the primary $\nu=1$ resonance due to the first pulse is very distorted by the
resonance.
Detailed calculation shows that there are small higher order
(non-primary) resonance islands between $J=3$ and $J=4$.  Fig. 2.d
shows the classical
phase space at time, $t_{fix}=t_s$ when the second pulse has reached its
maximum amplitude with $U_f(t_s)=0.055$ and $U_s(t_s)=3.0$.
         The primary $\nu=1$ resonance due to the first pulse is very
small and its region
of influence does not extend very high in energy.

It is interesting to compare the classical phase space for $U_o=3.0$
with a case when the maximum pulse amplitude is $U_o=0.5$.
In Figure 3, we show the strobe plots of the classical phase space at
the same times, $t_{fix}=t_1,~t_f,~t_c,~t_s,~t_2$  as in Fig. 2 but
with $U_o=0.5$.
More island structures survive with this relatively weak value of pulse
amplitude than in Fig. 2.
The invariant surfaces between the two $\nu=1$ primary resonances are
distorted and
higher order non-primary resonance islands can be seen even when the
pulse amplitudes have reached their maximum values.

\section {The Quantum system}

The Schr\"{o}dinger equation for the driven square-well system
described in Sect. 2 can be written (in dimensionless units)
\begin{equation}
i \frac{\partial}{\partial t} \langle x |\psi(t) \rangle = \left (
-\frac{\partial^2}
{\partial x^2} + U_f(t) x {\rm cos}(\omega_f t) + U_s(t) x {\rm cos}
(\omega_s t) \right ) \langle x |\psi(t) \rangle,
\label{schrodeq}
\end{equation}
where the momentum operator is given by ${\hat p}=-i\partial /\partial x$.
In order to satisfy the boundary condition at $x=\pm 1$, the wavefunction
should satisfy
$\psi(x={\pm}1,t)={\langle}x={\pm}1|\psi(t){\rangle}=0$ for all
times, $t$.

For the case of a quantum particle in the infinite square well, when
$U_{f,s}=0$ (no driving force), the energy is conserved and we have a
complete set of orthonormal energy eigenstates which can be used as
basis states to describe
the dynamics of the driven system.
For the unperturbed system, the energy eigenvalues are $E_n=n^2
\pi^2/4$ and the orthonormal energy eigenstates are
${\langle}x|E_n{\rangle}=\phi_n(x)={\rm sin}[ n\pi (x-1)/2]$.
The  dipole matrix elements in this basis are
$x_{n,n'}={\langle}E_n|{\hat x}|E_{n'}{\rangle}$
where
\begin{equation}
x_{n,n'} = \left \{ \begin{array}{ll}
0,  &[n +n'] ~~~~~(\mbox{modulo} ~~~ 2)=0 \\
\frac{16 n n'} {\pi^2 (n^2-{n'}^2)^2},
           &[n +n'] ~~~~~(\mbox{modulo} ~~~ 2)=1.
\end{array}  \right.
\end{equation}
Note that integer values of $J$ ($J=n$) in the classical Hamiltonian
correspond to the allowed quantized states of the quantum system.
This simplifies comparison between the classical and quantum systems.

We can expand the wavefunction, $|\psi(t){\rangle}$, in the
unperturbed energy basis so $|\psi(t){\rangle}=\sum_n c_n(t) |E_n
\rangle$. Then we can  rewrite the Schr\"{o}dinger equation in the form
\begin{equation}
\frac{d c_n(t)}{d t} = - i E_n c_n(t) + i [U_f (t) \mbox {cos}
(\omega_f t)  + U_s (t) \mbox {cos} (\omega_s t)]
        \sum_{n'} x_{n,n'} c_{n'}(t)
\label {schrodingerE}
\end{equation}
where $c_n(t)={\langle}E_n|\psi(t)\rangle$ is the probability
amplitude to find the system in
the $n$th energy level  at time $t$.
We will generally assume that at time $t=0$ the system is in
the state $|\psi(0){\rangle}=|E_1{\rangle}$.
We will then find the state $|\psi(+\infty){\rangle}$ after the
two pulses have been turned on and  off.

\section {Floquet States}

Once we fix the amplitudes, $U_{f}(t=t_{fix})=U_f(t_{fix})$ and
$U_{s}(t=t_{fix})=U_s(t_{fix})$,  the
Hamiltonian becomes time periodic and the Schr\"{o}dinger equation
takes the form
\begin{equation}
i \frac{\partial}{\partial t} \langle x |\psi(t) \rangle = \left (
-\frac{\partial^2}
{\partial x^2} + U_f(t_{fix}) x {\rm cos}(\omega_f t) + U_s(t_{fix})
x {\rm cos}
(\omega_s t) \right ) \langle x |\psi(t) \rangle.
\label{floqschod}
\end{equation}
For such systems, the energy is not conserved.
However, if the carrier frequencies of the pulses are commensurate so
${\omega_f/ \omega_s}={n_f/ n_s}$, where $n_f$ and $n_s$ are
integers, then the Hamiltonian
is invariant under a discrete time translation $H(t)=H(t+T_o)$, where
the period $T_o$ of the Hamiltonian  is
\begin{equation}
T_o=\pi{\biggl(}{{n_f \over \omega_f}+{n_s \over \omega_s}}{\biggr)}.
\label{period}
\end{equation}
For such systems, Floquet eigenstates, $|\phi_{\alpha}(t){\rangle}$
(which have period $T_o$ so
$|\phi_{\alpha}(t+T_o){\rangle}=|\phi_{\alpha}(t){\rangle}$)  form a
complete orthonormal basis which determines the dynamics.
Furthermore, the Floquet eigenphases, $\Omega_{\alpha}$, are conserved
quantities \cite{shirley}, \cite{sambe}, \cite{reichl}.

We can obtain an eigenvalue equation relating  $\Omega_{\alpha}$ and
$|\phi_{\alpha}(t)\rangle$.
Consider the case when the system is in the
$\alpha$th Floquet eigenstate  so that
$|\psi(t){\rangle}={\rm
e}^{-i\Omega_{\alpha}t}|\phi_{\alpha}(t){\rangle}$. Then substitution
into Eq. (\ref{floqschod})  yields the eigenvalue equation
\begin{equation}
{\biggl(}{\hat H}(t)-i{{\partial}\over
{\partial}t}{\biggr)}|\phi_{\alpha}(t){\rangle}=\Omega_{\alpha}|\phi_{\alpha}(t){\rangle},
\label{floqeigprob}
\end{equation}
where ${\hat H}_F(t){\equiv}{\hat H}(t)-i{{\partial}/ {\partial}t}$ is the
Floquet Hamiltonian.

More generally, when the system is in the state $|\psi(0)\rangle$ at
time $t=0$,
     the state of the system at time $t$ can be written
\begin{equation}
|\psi(t){\rangle}=\sum_{\alpha}A_{\alpha}{\rm
e}^{-i\Omega_{\alpha}t}|\phi_{\alpha}(t){\rangle}
=\sum_{\alpha}{\langle}\phi_{\alpha}(0)|\psi(0){\rangle}{\rm
e}^{-i\Omega_{\alpha}t}|\phi_{\alpha}(t){\rangle},
\end{equation}
The state of the system at time $t=T_o$ takes on an especially simple form
\begin{equation}
|\psi(T_o){\rangle}={\hat {\rm U}}_F(T_o)|\psi(0){\rangle}=\sum_{\alpha}{\rm
e}^{-i\Omega_{\alpha}T_o}|\phi_{\alpha}(0){\rangle}
{\langle}\phi_{\alpha}(0)
|\psi(0){\rangle}.
\end{equation}
where we have used the fact that
$|\phi_{\alpha}(T_o){\rangle}=|\phi_{\alpha}(0){\rangle}$.
The Floquet evolution operator, ${\hat {\rm U}}_F(T_o)$, can now be defined
\begin{equation}
{\hat {\rm U}}_F(T_o)=\sum_{\alpha}{\rm
e}^{-i\Omega_{\alpha}T_o}|\phi_{\alpha}(0){\rangle}{\langle}\phi_{\alpha}(0)|.
\end{equation}
We can compute matrix elements of the Floquet evolution operator in the basis
of unperturbed energy eigenstates.
Then the $(n,n')$th matrix element of the resulting
Floquet matrix is given by
\begin{equation}
U_{n,n'}(T_o)={\langle}E_n|{\hat {\rm
U}}_F(T_o)|E_{n'}{\rangle}=\sum_{\alpha}{\rm
e}^{-i\Omega_{\alpha}T_o}{\langle}E_n|\phi_{\alpha}(0){\rangle}{\langle}\phi_{\alpha}(0)|E_{n'}{\rangle}.
\end{equation}
The $\alpha$th eigenvalue of the Floquet matrix $U_{n,n'}(T_o)$ is ${\rm
exp}({-i\Omega_{\alpha}T_o})$ and the $\alpha$th eigenvector
        in the unperturbed energy basis is
given by a column matrix composed of matrix elements,
${\langle}E_n|\phi_{\alpha}(0){\rangle}$, where $n=1,...,\infty$. The
eigenvalues
$\Omega_{\alpha}$ can be obtained from ${\rm
exp}({-i\Omega_{\alpha}T_o})$, but only modulus $\omega_o$. We
refer to the eigenvalues
$\Omega_{\alpha}$ obtained from the Floquet matrix as {\it
eigenphases}.

For the system we consider here, the Floquet matrix has a natural
truncation which is
determined by the nonlinear dynamics of the system.
     Classically, the driven square-well
system has a region of mixed phase space bounded at high energies by
KAM tori.  For the cases we will consider here, where the initial state
$|\psi(0){\rangle}$ is the unperturbed energy level
$|\psi(0){\rangle}=|E_1{\rangle}$, the state $|\psi(0){\rangle}$ can never
penetrate very far into the high energy  KAM region. This
provides a natural truncation of the size of the Floquet matrix and
we need to include only enough unperturbed basis
states, $|E_n{\rangle}$, to cover adequately the region of mixed
phase space.  Each column of the Floquet matrix
can be constructed by solving the  time-dependent Schr\"{o}dinger
equation for one period, $T_o$, with the system
initially in one of the unperturbed energy eigenstates. This
integration is performed using  each of the unperturbed
energy eigenstates as an initial state until all the columns of the
Floquet matrix have been computed.
Floquet eigenphases and eigenstates are obtained by numerically diagonalizing
the Floquet matrix \cite{reichl}.

Husimi distributions allow us to visualize the distribution of
probability of the Floquet eigenstates in
the underlying classical phase space \cite{husimi}. Physically they
describe the location
of the particle in the presence of the radiation field and provide
important information
about the actual dynamics taking place in the system. The Husimi
distribution for a
Floquet eigenstate
$|\phi_{\alpha}
\rangle $ is defined as
$H(x_0,p_{0}) = | \langle \phi_{\alpha} |x_0,p_{0} \rangle |^2 $,
where the state $|x_0,p_{0} \rangle $ is a coherent state that can be
represented
in the position basis as \cite{chism}
\begin{equation}
\langle x | x_0, p_0 \rangle = \left (\frac {1}{\sigma^2 \pi} \right)^{1/4}
{\rm exp} \left(-\frac{(x-x_0)^2} {\sigma^2} + \frac {i p_0 (x - x_0)}
{\hbar}\right).
\end{equation}
The coherent state is a minimum uncertainty wave packet and has a parameter,
$\sigma$, that determines the relative dispersion in both position and
momentum space.

In the subsequent sections, we will consider three different choices
for carrier
frequencies of the
pairs of pulses which drive the system. For Case I,  the first pulse
connects levels
$n=2$ and $n=3$ and the second pulse connects levels $n=1$ and $n=2$.
This is the
traditional model for the STIRAP ladder process \cite{shoreA}.
However, as distinct from the usual
       discussion of STIRAP we will deal with
the exact dynamics of the system. We will take  account of the fact
that we have a
multilevel system that can undergo a
transition to chaos.  We will examine the effect of the full
nonlinear dynamics on this system.  For Case II,
the first pulse connects levels $n=4$ and $n=5$ and the second pulse
connects levels $n=1$ and $n=4$. This is again a ladder process. For
this case the
underlying chaotic dynamics will have a surprising effect on the
transition probabilities.
Finally, for Case III, we consider a {\it lambda} process in which
the first pulse
connects levels  $n=3$ and $n=4$ and the second pulse
connects levels $n=1$ and $n=4$. In all cases, we consider the {\it
exact} dynamics of
the driven system.

\section {Case I: First pulse $2{\rightarrow}3$; Second pulse
$1{\rightarrow}2$}

In this section, we examine the dynamics of the driven square-well
system when two pulses
are applied such that the first pulse connects levels $n=2$ and $n=3$ and
the second pulse then connects the
levels $n=1$ and $n=2$. We first determine the behavior of Floquet
eigenstates at fixed times $t=t_{fix}$
during which the pulses drive the system. The distribution of
probability in the Floquet
eigenstates is sensitive to structures in the classical phase space
which are larger
than Planck's constant. We then compare the prediction of Floquet
theory to the actual
behavior of the system in the non-adiabatic  and adiabatic regimes.

\subsection{Behavior of Floquet Eigenstates}

The first  pulse has carrier frequency
${\omega}_f=(E_3-E_2)=5 \pi^2/4$ and the
second pulse has carrier frequency
$\omega_s=(E_2-E_1)=3 \pi^2/4$. These  frequencies
are commensurate since ${\omega_f/ \omega_s}={5/ 3}$.   From
Eq. (\ref{period}), the period of
the Hamiltonian is $T_o=8/\pi$ and the {\it Floquet frequency} is
${\omega}_o=2{\pi}/T_o=\pi^2/4$. Thus, ${\omega}_f=5{\omega}_o$ and
${\omega}_s=3{\omega}_o$. We  set $U_o=3.0$.

The dynamics of this system tells us that we only need to keep five
unperturbed energy
eigenstates as a basis to form the Floquet matrix.   This can  be
seen from Fig. 2
where we show the underlying classical phase space at selected values of
$t_{fix}$ during the time that the pulses are on.  For $J>5$, the
classical phase
space is dominated by KAM tori with almost constant values of $J$ and
the unperturbed
energy states are very weakly coupled by the dynamics for
$n>5$.  Thus, to describe the quantum behavior of this system, it is
sufficient to
construct a $5{\times}5$ Floquet matrix with the five basis states
$|E_1{\rangle},...,|E_5{\rangle}$. We find that only four of the five
eigenstates of
the Floquet matrix are actively involved in the dynamics.  Their eigenphases,
$\Omega_{\alpha}$, are plotted modulo
${\omega}_o=\pi^2/4$ in Figure 4.a.  Two of these Floquet eigenphases
are almost degenerate over the time interval
that the  pulses act and are not distinguishable on the scale shown in Fig.
4.a.  For $t_{fix}=0$, the four Floquet eigenphases are approximately
degenerate modulo ${\omega}_o$.

We can follow each Floquet eigenstate during the entire process by
computing the eigenstates for a sequence of values of $t_{fix}$
over the interval $0{\leq}t_{fix}{\leq}t_{tot}$. For closely spaced
values of $t_{fix}$, Floquet eigenstates at different times belonging
to different
eigenphases will be orthogonal.
This provides a means of following
the evolution of each eigenstate as a function of $t_{fix}$.
As we will see,  the Floquet eigenstates can change structure when
avoided crossings occur between Floquet eigenphases. To keep track of the
changes that occur in the Floquet eigenstates,
we will give each eigenstate a unique alphabetical label determined
by its dominant dependence on unperturbed
energy states at time $t_{fix}=0$. We find that at $t_{fix}=0$ the Floquet
eigenstates have the following structure and we
give them the following labels:
\begin{eqnarray}
A=|\phi_1\rangle=|E_1\rangle,~
~~D=|\phi_4\rangle=|E_4\rangle ~{\rm and}~
~~E=|\phi_5\rangle=|E_5\rangle, \nonumber\\
BC^+=|\phi_2\rangle={1\over \sqrt{2}}(|E_2\rangle+|E_3\rangle),
~~~BC^-=|\phi_3\rangle={1\over \sqrt{2}}(|E_2\rangle-|E_3\rangle).
\end{eqnarray}

The traditional STIRAP ladder process assumes that the molecule or atom in
question can be approximated by  a three level
system and causes a coherent population shift of the
atom from level 1 to level 3.  We find that the traditional STIRAP
ladder process occurs in
our system for
$U_o<0.1$.  For amplitudes $U_o<0.1$,  state $D=|\phi_4\rangle$ does
not participate in the dynamics
at all. The Floquet eigenvalue curve for $\Omega_{4}$, plotted as a
function of $t_{fix}$, crosses that
for $\Omega_{1}$ in two places but does not undergo any avoided
crossings. The state
$D=|\phi_4\rangle$ remains predominantly dependent on $|E_4\rangle$
during the entire process.  For traditional
STIRAP, curve D in Fig. 4.b is replaced by state $A$ and the only avoided
crossing that occurs is the three-way avoided
crossing at $t_{fix}=t_c$ between states $A$, $BC^+$, and $BC^-$.
The state $A$, which is predominantly composed of the state
$|E_1\rangle$ before the multiple avoided crossing at $t=t_c$, becomes
predominately dependent on state $|E_3\rangle$ after the multiple
avoided crossing, having interchanged its ``1" character
with the  ``3" character of states $BC^+$ and $BC^-$ at the avoided
crossing. Thus, at the end of the process state
$A$ would be composed  predominately of level $n=3$ and the states
$BC^+$ and $BC^-$ would be predominately of
superpositions of levels $n=1$ and $n=2$.

Once the amplitude $U_o$ becomes greater than $U_o=0.1$, something
different happens due to the avoided crossing shown in Fig.
4.c.  For $U_o<0.1$ the Floquet eigenphases for states $A$ and $D$ in
Fig. 4.c {\it cross}
at time $t_{fix}=\tau_I{\approx}{10\over 23} t_{tot}$ just before
$t_{fix}=t_c$. For $U_o>0.1$ the Floquet
eigenphases for states $A$ and $D$ in Fig. 4.c  {\it avoid crossing}
at time $t_{fix}=\tau_I$.  Before time $t_{fix}=\tau_I$ Floquet state $A$ is
predominately composed of level
$n=1$ and Floquet state D is predominately composed of level
$n=4$. After the avoided crossing at time $t_{fix}=\tau_I$ the states
have changed their character and Floquet state $A$ is composed
predominately of level $n=4$ and Floquet state $D$ is composed
predominately of level $n=1$.
Because of the avoided crossing at $t_{fix}=\tau_I$, the
entire population gets shifted from level $n=1$ to level $n=4$ {\it before}
the traditional STIRAP ladder process can take place. The traditional 
STIRAP ladder process now occurs among {\it unpopulated}  states.
It is interesting to note that the Floquet states $A$ and $D$  {\it
cross} at time  $t_{fix}=\tau_{II}{\approx}{13/23}t_{tot}$. A symmetry that
was broken earlier appears to have been restored.
  These transitions are clearly seen in Fig. 5 where
we show the level compositions of the four participating Floquet
states, $A$, $BC^+$, $BC^-$ and $D$
as a function of $t_{fix}$.

\subsection {The Population Transfer}

   Let us now determine the exact behavior of the system,
when the pulses are
applied, by solving the Schr\"{o}dinger equation in Eq.
\ref{schrodingerE}.  We will assume that at time $t=0$ the system is in state
$|\psi(0){\rangle}=|E_1{\rangle}$.
As we will see, the actual dynamics of this system is determined by
the length of time during which the pulses are allowed
to act.
The pulse duration time necessary to achieve adiabatic behavior of the
system is determined largely by the
avoided crossings in the Floquet eigenphases. At isolated avoided
crossings, involving only two eigenstates, the states
involved interchange their character.

Avoided crossings of Floquet
eigenphases occur as the classical phase space becomes chaotic, and a
symmetry has been
broken in that local region of the phase space.
The probability $P_{LZ}$ that a transition occurs between the two
Floquet eigenstates involved in an {\it isolated} avoided crossing
can be computed from a formula
obtained independently by Landau \cite{landau} and Zener \cite{zener}.
  For
our system, the Landau-Zener probability is given by
\begin{equation}
P_{LZ} = \mbox {exp}\left( -\frac{\pi (\delta \epsilon)^2}{2 \gamma}\right),
\end{equation}
where $\delta \epsilon$ is the eigenphase spacing at the avoided crossing
and $\gamma$ is the rate
of change of the Floquet eigenphases with respect to time  $t_{fix}$
in the neighborhood
of  the avoided crossing.

We have computed the Landau-Zener probability $P_{LZ}$ for the isolated
sharp avoided crossing at time $t_{fix}=\tau_I$
shown in Fig. 4.c. The Landau-Zener probability  depends on
$t_{tot}$. The larger
$t_{tot}$, the more ``stretched out" the horizontal axis in Fig. 4.c
will be relative to the vertical axis.  We have obtained the
following results by analyzing
Fig. 4.c for different values of $t_{tot}$.  For $t_{tot}=120$,
$\delta \epsilon=0.005$, and $\gamma=0.001875$
giving a Landau-Zener probability $P_{LZ}=0.979270$.  For
$t_{tot}=21000$, $\delta \epsilon=0.0063$, and
$\gamma=0.00001376$ giving a Landau-Zener probability
$P_{LZ}=0.0108$.  For $t_{tot}=270000$, $\delta
\epsilon=0.0030$, and $\gamma=1.2{\times}10^{-7}$ giving a
Landau-Zener probability $P_{LZ}{\approx}0$. The first case is
not in the adiabatic regime, but the second two cases are in the
adiabatic regime because the probability of a transition is negligible.

In Fig. 6, we show the probability
$P_n(t)=|{\langle}E_n|\psi(t){\rangle}|^2$ (for the four levels
$n=1,~2,~3,~4$) to find
the system in the $n$th unperturbed level at time $t$ for the three
cases; $t_{tot}=120$,
       $t_{tot}=21000$ and  $t_{tot}=270000$. These results are obtained by
directly solving the
Schr\"odinger equation, Eq. (\ref{schrodingerE}).  In all cases we start the
system in the initial state,
$|\psi(0)\rangle=|E_1\rangle$  with maximum pulse strength,
$U_o=3.0$.  In Fig. 6.a, where there is a large
Landau-Zener probability for the system to jump from Floquet state
$A$ to Floquet state $D$, the system comes out of the sharp
avoided crossing at $t_{fix}=\tau_I$ still predominately in the level
$|E_1\rangle$, and
the traditional  STIRAP ladder process can then occur at $t=t_c$. As
the pulses are
turned on and off, the system transitions  from level
$|E_1\rangle$ to level $|E_3\rangle$. In Figs. 6.b and 6.c, the
Landau-Zener probability
is essentially zero and no transition occurs at
the sharp avoided crossing at $t_{fix}=\tau_I$. The system comes out
of the sharp avoided
crossing in level $|E_4\rangle$.  As the laser pulses are turned on
and off the system
transitions from the initial state
$|\psi(0)\rangle=|E_1\rangle$ to the final state
$|\psi(+\infty)\rangle=|E_4\rangle$.   Note that  both Figs. 6.b and 6.c
follow almost exactly the behavior of the Floquet state
$A$ shown in Fig. 5.c. This is an indication that we are
in the adiabatic regime  in Figs. 6.b and 6.c.

The  very large oscillations in the probability in Figs. 6.b and 6.c
have  been explained
by Berry \cite{berry} in terms of a sequence of "super-adiabatic
bases". He shows that the
decrease in the amplitude of these oscillations  as we increase
$t_{tot}$ is a sign that we are moving further into the adiabatic regime.  The
frequencies of the oscillations in Figs. 6.b and 6.c appear to be
determined by the
difference in Floquet eigenphases of the two Floquet states
involved in the sharp avoided crossing. For example, at $t_{fix}=t_f$ the
period of the oscillation is
$T_{osc}{\approx}400$. The difference in the Floquet eigenphases is
$|\Delta\Omega|=|{\Omega}_1-{\Omega}_4|{\approx}0.016$. Thus,
$T_{osc}=2\pi/|\Delta\Omega|=393$.  Similarly, at $t_{fix}=(\tau_I-t_f)/2$,
$T_{osc}{\approx}600$. The
difference in the Floquet eigenphases is
$|\Delta\Omega|=|{\Omega}_1-{\Omega}_4|{\approx}0.011$. Thus,
$T_{osc}=2\pi/|\Delta\Omega|=571$. The observed oscillation periods
are the same for both Fig. 6.b and Fig. 6.c.

\section {Case II: First pulse $4{\rightarrow}5$; Second pulse
$1{\rightarrow}4$}

We now turn on pulses with higher carrier  frequencies in order to
examine more closely the relation between the
quantum transitions
and their  relation to the underlying classical dynamics.
We first apply a pulse whose carrier frequency is
${\omega}_f=(E_5-E_4)=9 \pi^2/4$. We then apply a
second pulse whose carrier frequency is  $\omega_s=(E_4-E_1)=15
\pi^2/4$. The two frequencies
are commensurate since ${\omega_f/ \omega_s}={3/ 5}$.
       From Eq. (\ref{period}), the period
of the Hamiltonian and the Floquet frequency  are again $T_o=8/\pi$
and ${\omega}_o=\pi^2/4$, respectively. Thus,
${\omega}_f=9{\omega}_o$ and
${\omega}_s=15{\omega}_o$. We will  consider the case when the
maximum amplitude of both
pulses is
$U_{o}=13.0$. For these frequencies and
amplitudes, we find that we can induce a transition of the entire population
from level $|E_1\rangle$ to level $|E_{10}\rangle$. Below we describe
how this happens.

   Classical phase space plots for times $t_{fix}=t_f$,
$t_{fix}=t_c$, $t_{fix}=\tau_{IV}={3/5}t_{tot}$, and
$t_{fix}=t_s$  are shown in Figs. 7.a-7.d, respectively.
The first primary resonance due to the first pulse is located at $J=4.5$  and
the first primary resonance due to the second pulse is located at $J=7.5$.
The frequency of the second pulse is chosen to connect levels $n=1$ and
$n=4$. However it
also connects the levels $n=7$ and $n=8$ since
$E_{8}-E_7=15{\omega}_o$.
This is why the first primary resonance due to the second pulse lies
at $J=7.5$.
The states below $J=4.0$ are immersed in the chaotic sea formed by
the higher order primaries induced by the two pulses
during most of the time that the one or the other of the pulses have
a significant strength.
Higher-order non-primary resonance islands can be seen above $J=6.0$ during
most of the pulse sequence.
In Fig. 7.b there is a chaotic sea which has formed throughout the
region $J=0$ to $J=9$.
Also, during the last half of the pulse sequence, the KAM tori near $J=10$
are highly distorted due to the formation of the primary resonance at $J=7.5$.
Thus, from the classical phase space we see that it requires
approximately twelve
square-well energy eigenstates to accurately describe the dynamics of
this system.

The Floquet  matrix that we use to describe the quantum dynamics  is
a $12{\times}12$
matrix. However,  we find that only ten Floquet eigenstates are
directly involved in the
dynamics. To keep track of these ten Floquet eigenstates,  we
will give
each state a unique alphabetical label determined by their dominant
dependence on unperturbed energy states at time
$t_{fix}=0$. We find that at $t_{fix}=0$ the Floquet eigenstates have the
following structure and we give them the following
labels:
\begin{eqnarray}
A=|\phi_1\rangle=|E_1\rangle,~~~B=|\phi_2\rangle=|E_2\rangle,~~~
C=|\phi_3\rangle=|E_3\rangle,~~~F=|\phi_6\rangle=|E_6\rangle\nonumber\\
DE^+=|\phi_4\rangle={1\over \sqrt{2}}(|E_4\rangle+|E_5\rangle),
~~~DE^-=|\phi_5\rangle={1\over \sqrt{2}}(|E_4\rangle-|E_5\rangle), \nonumber\\
G=|\phi_7\rangle=|E_7\rangle,~ ~~H=|\phi_8\rangle=|E_8\rangle,~ ~~
I=|\phi_9\rangle=|E_9\rangle,~~~J=|\phi_{10}\rangle=|E_{10}\rangle.\nonumber\\
\label{order}
\end{eqnarray}
The Floquet eigenphases corresponding to these ten Floquet eigenstates
       are plotted modulo $\omega_o$ in Fig. 8.
A number of  avoided crossings  occur between the
eigenphases during the time the pulses act on the system. There are
four  avoided crossings that largely determine
the dynamics.  There is a multiple wide avoided crossing at
$t_{fix}=t_c={1/\over 2}t_{tot}$ which involves the  seven
states, $B$, $C$, $DE^{\pm}$, $F$,  $H$, and $I$. There is a sharp
avoided crossing at
$t_{fix}=\tau_{III}{\approx}{3\over 8}t_{tot}$ that involves the states
$B$ and $G$.  There is a three-state wide
avoided crossing at
$t_{fix}=\tau_{IV}{\approx}{3\over 5}t_{tot}$ that involves states
$A$, $B$ and $H$.  There is a sharp avoided
crossing at
$t_{fix}=\tau_{V}{\approx}{2\over 3}t_{tot}$ which involves the
states $A$ and $J$.

In Fig. 9.a we plot the eigenphases of the seven Floquet
states $B$, $C$,
$DE^{\pm}$, $F$, $H$,
and $I$ involved in the multiple avoided
crossing at
$t_{fix}=t_c={1\over  2}t_{tot}$.  These states have support in the
unperturbed square-well levels $E_3$, $E_4$, $E_5$, $E_6$, $E_7$,
$E_8$, and $E_9$.  In Fig. 9.b we show a magnification
of the very sharp avoided crossing between
states $DE^-$ and $H$ at $t_{fix}=t_c={1\over 2}t_{tot}$.  The effect of
these avoided
crossings can be seen in the dependence of the Floquet eigenstates on the
square-well states $|E_n{\rangle}$. Plots of Floquet eigenstates $B$, $H$, and
$G$ are shown in Fig. 10 and plots of Floquet eigenstates $DE^{\pm}$,
$C$ and $F$ are
shown in Fig. 11. There is a complicated interchange of levels
occurring. As shown in
   \cite{timber}, at such multiple avoided crossings, the Floquet 
eigenstates emerge with
   very different probability distributions than the entering states.
Thus, multiple avoided crossings behave differently from isolated 
pairs of avoided
crossings where the
states simply interchange character.  Multiple avoided crossings
provide a mechanism for
the spread of the manifestations of chaos in quantum systems   \cite{timber}.

It is useful to note that there is
an  isolated avoided crossing at $t_{fix}=\tau_{III}$
that causes states $B$ and $G$ to switch from
$B{\approx}|E_2\rangle$ and $G{\approx}|E_7\rangle$, as they enter the
avoided crossing at ${\tau}_{III}$, to
$B{\approx}|E_7\rangle$ and
$G{\approx}|E_2\rangle$ as they leave.
Thus state $B{\approx}|E_7\rangle$ as it enters the multiple avoided crossing
at   $t_{fix}=t_c={1\over 2}t_{tot}$.

Let us now consider the transition that causes the population of the
square-well to undergo a coherent transition
from level $|E_1\rangle$ to level $|E_{10}\rangle$. This can occur if
the system evolves adiabatically and follows
the behavior of Floquet state $A$. The level dependence of Floquet
state $A$ is shown in Fig. 12.a. It starts out in
level $|E_1\rangle$ and then partially switches to level
$|E_5\rangle$ at $t_{fix}=\tau_{IV}$
due to a three state avoided crossing between states $A$,
$B$, and $H$, and finally at
$t_{fix}=\tau_{V}$ it switches completely
to level $|E_{10}\rangle$ due to a sharp
avoided crossing between states $A$ and $J$.  The avoided crossings
at times $t_{fix}=\tau_{IV}$ and
$t_{fix}=\tau_{V}$ that
cause these transitions are shown
in Figs. 13.a and 13.b, respectively.

In Fig. 14 we show a sequence of Husimi plots of the states $B$, $H$,
$A$, and $J$ as they go through the avoided
crossings at $t_{fix}=\tau_{IV}$
and $t_{fix}=\tau_{V}$.
The Husimi plots show the location of the quantum
particle in phase space when the system is in a given eigenstate. The 
columns  from left to right, show the
Floquet states  $B$, $H$, $A$ and $J$.    From top to bottom,
the Husimi plot of each state is shown at times (a) $t_{fix}=t_f$,
(b) $t_{fix}=t_c$, (c) $t_{fix}=\tau_{IV}$, (d) $t_{fix}=\tau_{V}$ and (e)
$t_{fix}=t_2$.  The sequence of events causing Floquet state $A$ to
undergo a transition from level $|E_1\rangle$ to level
$|E_{10}\rangle$ can be seen clearly in these plots.

Finally in Figs. 15.a and 15.b, we plot the probability
$P_n(t)=|{\langle}E_n|\psi(t){\rangle}|^2$ (for the ten levels
$n=1,...,10$) to find the system in the $n$th unperturbed level at
time $t$ (Fig. 15.c will be discussed in Sect. 7).  The system is 
prepared in initial state $|\psi(0)\rangle=|E_1\rangle$
  with maximum pulse strength, $U_o=13.0$.  We obtain these plots by directly
solving the Schr\"{o}dinger equation in Eq. (\ref{schrodingerE}).
In Fig. 15.a we show the results for the nonadiabatic case. We choose
$t_{tot}=600$ which is a rapid evolution of the pulses. After the
pulses have passed, the final state of the system is
$|\psi(+\infty)\rangle=|E_5\rangle$. The adiabatic case is shown in Fig.
15.b. We now choose $t_{tot}=6000$. The
probabilities now closely follow the behavior of the single Floquet
eigenstate $A$ as can be seen in Fig. 12.a.
After the pulses have passed, the final state of the system is
$|\psi(+\infty)\rangle=|E_{10}\rangle$.
This  transition dynamics
is determined by the structure of the avoided crossings in the Floquet
eigenphase curves and by the phase space structure of the Floquet
eigenstates at the avoided crossing.

The phase space structure of the Floquet eigenstates (see the
Husimi plots) is determined by  structures in the underlying
classical phase space which
are greater than Planck's constant. The transition from
$|\psi(0)\rangle=|E_1\rangle$ to
$|\psi(\infty)\rangle=|E_{10}\rangle$ occurs because at the time of
the transition the classical phase space is connected by a chaotic
sea from $J=1$ to $J=9$ and a
strongly distorted KAM region at $J=10$ which allows the quantum
state to tunnel into the
phase space region around $J=10$. This can be seen from the Husimi
plot for the state
$A$  at time $t_{fix}={13\over 20} t_{tot}$ which is shown in Fig. 16.
The unperturbed levels, $n=1$, $n=5$ and $n=10$ are connected via the
chaotic sea around the two $\nu=1$ primary resonances
which exist at the time
$t_{fix}={13\over 20} t_{tot}$.

\section {Case III: First pulse $3{\rightarrow}4$; Second pulse
$1{\rightarrow}4$}

The final case we consider is a  lambda process in which
the system is driven by pulses with carrier frequencies
$\omega_f=(E_4-E_3)=7{\pi}^2/4$ and $\omega_s=(E_4-E_1)=15{\pi}^2/4$.
These frequencies
are commensurate and the period of the Hamiltonian and the Floquet
frequency are again
$T_o=8/\pi$ and ${\omega}_o={\pi}^2/4$, respectively. To stay as
close to Case II
parameters as possible, we chose pulse amplitudes $U_o=13.0$.
Although this case looks
similar to Case II, we will find quite different results.

In the adiabatic limit, the
transition $|E_1{\rangle}{\rightarrow}|E_3{\rangle}$ occurs, as shown
in Fig. 15.c, which
is very different from Case II.
We believe that the difference between these two processes is due to
the difference in
the  phase space distribution of the resonances and chaos induced in
the system by the
laser pulses.  Strobe plots of the classical phase space for the lambda case
for times $t_{fix}=t_f$, $t_{fix}=t_c$, $t_{fix}=\tau_{IV}={0.62}t_{tot}$, and
$t_{fix}=t_s$  are shown in Figs. 17.a-17.d, respectively.
The first primary resonance due to the first pulse is located at
$J=3.5$  (rather than
$J=4.5$ as is in Case II) and the first primary resonance due to the
second pulse is
located at $J=7.5$ (the same as Case II). For Case II,  a pathway is
opened by a chaotic
sea that allows a Floquet state to tunnel across the entire energy
region (see Fig. 16)
from $|E_1\rangle$ to $|E_{10}\rangle$ at the time of the avoided crossing.
In the strobe plots of the classical phase space in Fig. 17 for Case
III, this pathway is
blocked by KAM tori which  do not allow the quantum system to tunnel
out of the low
energy region. We also find fewer avoided crossings in the plot of the Floquet
eigenphases for this case.

\section {Conclusion}

We have found that sequential laser pulses, when applied to a
particle in an anharmonic multilevel system, can induce nonlinear
resonances and a
transition to chaos in the  dynamics of the particle. The extent of
the region influenced
by resonances and chaos determines the number of unperturbed energy
eigenstates that must
be kept to form the basis used to construct the Floquet matrix and to
determine the
quantum dynamics.

When the
pulses are applied in a manner which adiabatically changes the
dynamics of the particle, this
transition to chaos can be used to control the coherent transfer of
the particle across the chaotic
sea from a low lying energy state to a highly excited energy state.
Floquet theory provides an
accurate means of describing the dynamical behavior of these driven
systems in the adiabatic limit.
The type of population transfer that is allowed depends on the nature
of the Floquet
eigenphase avoided crossings created by the underlying transition to
chaos and the ability of the corresponding Floquet eigenstates to
tunnel across vast
regions of the phase space because of the induced chaos. In the Cases I and II,
avoided crossings occurred between Floquet eigenstates which had
spread throughout the
available phase space, and a  coherent flow of probability from the
low energy  side
   to the high energy side of the chaotic sea occurred. In Case III,
the path to the high
energy states around the resonance at $J=7.5$ appears to be blocked
by KAM tori and the
transition to the higher energy states does not occur.

In molecular systems, the success of STIRAP will depend on the state
of the dynamics of
the unperturbed molecular system, and on the changes in that dynamics
induced by the
laser pulses.  Most molecular systems have regimes of internal chaos
and the interplay of
these regimes with the dynamics induced by the laser pulses is
critical to understanding
STIRAP in these system.

\section {Acknowledgments}

The authors wish to thank the U.S. Navy Office of Naval Research (Grant No.
N00014-03-1-0639) for support of this work and we wish to thank the Engineering
Research Program of the Office of
Basic Energy Sciences at the U.S. Department of Energy (Grant No.
DE-FG03-94ER14465) for partial
support of this work. Author L.E.R wishes to thank the Robert A.
Welch Foundation (Grant No. F-1051)
for partial support of this work. Both authors thank Dario Martinez
for the useful discussions about
Floquet theory. We also thank the University of Texas High
Performance Computing Center for
use of its facilities.
%
%


\pagebreak

\begin{figure}
\caption{Schematic diagram for the two pulses. The first pulse
connecting levels $|E_2\rangle$ and $|E_3\rangle$ is shown as a solid line.
The second pulse connecting levels $|E_1\rangle$ and $|E_2\rangle$ is shown
as a dotted line. They have
        maximum strength $U_o$ at times $t=t_f$ and $t=t_s$, respectively.
The whole pulse sequence takes a time $t=t_{tot}$ to complete.  In
the figure, $t_1=
{1/ 20}t_{tot}$, $t_c={1/ 2}t_{tot}$  and $t_2={19/ 20}t_{tot}$.}
\label{fig.1}
\end{figure}

\begin{figure}
\caption{Strobe plots of the action-angle variables $(J,\theta)$ for
the infinite square well system with pulse amplitudes $U_o=3.0$ and frequencies
$\omega_f=5\omega_o$ and $\omega_s=3\omega_o$.  Strobe plots are
shown at times (a)
$t_{fix}=t_1$, (b) $t_{fix}=t_f$, (c)
$t_{fix}=t_c$, (d)
$t_{fix}=t_s$ and (e) $t_{fix}=t_2$. For each plot $0{\leq}\theta{\leq}\pi$.
The three largest primary resonances $\nu=1,~2,~3$ due to the first
pulse are located at
$J=2.5, 0.83$ and $0.5$.
The three largest primary resonances$\nu=1,~2,~3$ due to the second
pulse are located at
$J=1.5, 0.5$ and $0.3$.}
\label{fig.2}
\end{figure}

\begin{figure}
\caption{Strobe plots of the action-angle variables $(J,\theta)$ for
the infinite square well system with pulse amplitudes $U_o=0.5$ and frequencies
$\omega_f=5\omega_o$ and $\omega_s=3\omega_o$,
respectively.  Strobe plots are shown at times (a)
$t_{fix}=t_1$, (b) $t_{fix}=t_f$, (c)
$t_{fix}=t_c$, (d)
$t_{fix}=t_s$ and (e) $t_{fix}=t_2$. For each plot $0{\leq}\theta{\leq}\pi$. }
\label{fig.3}
\end{figure}

\begin{figure}
\caption{Floquet eigenphases, for the system with maximum pulse strength
$U_o=3.0$  and frequencies
$\omega_f=5\omega_o$ and $\omega_s=3\omega_o$, are plotted over the
entire interval $0{\leq}t_{fix}{\leq}t_{tot}$.
(a) Floquet eigenphases for four Floquet states $A$, $BC^+$, $BC^-$ and
$D$  plotted mod
$\omega_o={\pi^2}/4$.  (b) Floquet eigenphase curves for the Floquet states
$D$, $BC^+$and  $BC^-$. The three-level wide avoided crossing at
$t_{fix}=t_c$ is clear.
(c) Floquet eigenphase curves for the Floquet states $A$ and $D$. The
sharp avoided crossing at
$t_{fix}=\tau_{I}$ and the crossing at $t_{fix}=\tau_{II}$ are clearly seen.}
\label{fig.4}
\end{figure}

\begin{figure}
\caption{Probability distribution
$|{\langle}E_n|\phi_{\alpha}{\rangle}|^2$ of the unperturbed energy
levels
$|E_n\rangle$ which compose each of the  Floquet eigenstates
(a) $BC^-$ (b) $BC^+$, (c) $A$ and (d) $D$,
plotted
       over the entire interval $0{\leq}t_{fix}{\leq}t_{tot}$ for pulse
amplitude $U_o=3.0$  and frequencies
$\omega_f=5\omega_o$ and $\omega_s=3\omega_o$.
The probability curve for level $|E_n\rangle$ is labeled with level
       quantum number $n$.}
\label{fig.5}
\end{figure}

\begin{figure}
\caption{The probability $P_n(t)=|{\langle}E_n|\psi(t){\rangle}|^2$ to find
the system in the unperturbed level
$|E_n\rangle$  for the system prepared in initial state
$|\psi(0)\rangle=|E_1\rangle$
       with maximum pulse strength, $U_o=3.0$  and frequencies
$\omega_f=5\omega_o$ and $\omega_s=3\omega_o$. The total
pulse duration times are  (a) $t_{tot}=120$, (b) $t_{tot}=21000$, and
(c)  $t_{tot}=270000$.
The numbers attached to each curve show the
components of the transition
probability in terms of the unperturbed energy eigenstate basis.
Case (a) is not in the adiabatic regime. Cases (b) and (c) are within
the adiabatic regime and basically reproduce
the structure of the single Floquet eigenstate $A$ in Fig. 5.c.}
\label{fig.6}
\end{figure}

\begin{figure}
\caption{Strobe plots of the action-angle variables $(J,\theta)$ for
the infinite square well system with
pulse  amplitudes $U_o=13.0$ and frequencies
$\omega_f=9\omega_o$ and $\omega_s=15\omega_o$,
respectively.  Strobe plots are shown at times (a) $t_{fix}=t_f$, (b)
$t_{fix}=t_c$, (c) $t_{fix}=\tau_{IV}={3/5}t_{tot}$ and
(d) $t_{fix}=t_s$.  The first primary resonance  from
the first pulse is located at
$J=4.5$  and the first primary resonance from the second pulse is located
at $J=7.5$.}
\label{fig.7}
\end{figure}

\begin{figure}
\caption{The ten Floquet eigenphases, plotted modulo $\omega_o$,
which determine the dynamics
for pulse strength $U_{o}=13.0$ and frequencies $\omega_f=9\omega_o$
and $\omega_s=15\omega_o$. The curves are
identified following the classification scheme in Eq. \ref{order} }
\label{fig.8}
\end{figure}

\begin{figure}
\caption{(a) The seven Floquet eigenphases, plotted modulo
$\omega_o$, which are involved in the multiple avoided
crossing at $t_{fix}=t_c={1\over 2}t_{tot}$ for pulse strength
$U_{o}=13.0$ and frequencies $\omega_f=9\omega_o$ and
$\omega_s=15\omega_o$. The curves are identified following the
classification scheme in Eq.
\ref{order}. (b) A magnification of the very sharp avoided crossing
between Floquet states $DE^-$ and $H$ at $t_{fix}=t_c={1\over 2}t_{tot}$.}
\label{fig.9}
\end{figure}

\begin{figure}
\caption{Probability distribution $|{\langle}E_n|\phi_{\alpha}{\rangle}|^2$
of the unperturbed energy levels which compose each of
the  Floquet eigenstates  (a) $B$, (b) $H$ and (c) $G$,  plotted
       over the entire interval $0{\leq}t_{fix}{\leq}t_{tot}$ for pulse
strength $U_o=13.0$ and frequencies
$\omega_f=9\omega_o$ and $\omega_s=15\omega_o$.  The probability
curve for level $|E_n\rangle$ is labeled with level
       quantum number $n$.}
\label{fig.10}
\end{figure}

\begin{figure}
\caption{Probability distribution $|{\langle}E_n|\phi_{\alpha}{\rangle}|^2$
of the unperturbed energy levels which compose each of
the  Floquet eigenstates  (a) $DE^-$, (b) $DE^+$, (c) $C$ and (d)
$F$  plotted
       over the entire interval $0{\leq}t_{fix}{\leq}t_{tot}$ for pulse
strength $U_o=13.0$ and frequencies
$\omega_f=9\omega_o$ and $\omega_s=15\omega_o$.  The probability
curve for level $|E_n\rangle$ is labeled with level
       quantum number $n$.}
\label{fig.11}
\end{figure}

\begin{figure}
\caption{Probability distribution $|{\langle}E_n|\phi_{\alpha}{\rangle}|^2$
of the unperturbed energy levels which compose each of
the  Floquet eigenstates  (a) $A$ and (b) $J$,  plotted
       over the entire interval $0{\leq}t_{fix}{\leq}t_{tot}$ for pulse
strength $U_o=13.0$ and frequencies
$\omega_f=9\omega_o$ and $\omega_s=15\omega_o$.  The probability
curve for level $|E_n\rangle$ is labeled with level
       quantum number $n$.}
\label{fig.12}
\end{figure}

\begin{figure}
\caption{(a) Magnification of the three Floquet eigenphases for
eigenstates $A$, $B$ and $H$
which are
involved in the three-state avoided
crossing at $t_{fix}=\tau_{IV}={3/5}t_{tot}$ for pulse
strength $U_{o}=13.0$ and frequencies
$\omega_f=9\omega_o$ and
$\omega_s=15\omega_o$. The curves are identified following the
classification scheme in Eq.
\ref{order}.  (b) Magnification of the
sharp avoided crossing at $t_{fix}=\tau_{V}={2/ 3}t_{tot}$
for pulse strength $U_{o}=13.0$ and
frequencies $\omega_f=9\omega_o$ and $\omega_s=15\omega_o$. Curves of
    eigenphases for Floquet eigenstates $A$
and $J$ are shown.}
\label{fig.13}
\end{figure}

\begin{figure}
\caption{Husimi plots for the Floquet states (a) $B$, (b) $H$, (c)
$A$ and (d) $J$
plotted in columns from left to right, respectively, for maximum
pulse   strength $U_o=13.0$ and frequencies
$\omega_f=9\omega_o$ and $\omega_s=15\omega_o$. Each Floquet state
(column) is shown (from top to bottom) at times
(a)
$t_{fix}=t_f$, (b) $t_{fix}=t_c$, (c)
$t_{fix}=\tau_{IV}={3/5}t_{tot}$,  (d)
$t_{fix}=\tau_{V}={2/3}t_{tot}$, (e)
$t_{fix}=t_2$. The effect of the avoided crossings that
enable Floquet state $A$ to undergo a
transition from level $|E_1\rangle$ to level $|E_{10}\rangle$ can be
seen clearly.}
\label{fig.14}
\end{figure}

\begin{figure}
\caption{The probability $|{\langle}E_n|\psi(t){\rangle}|^2$ to find
the system in the unperturbed level
$|E_n\rangle$  for the system prepared in initial state
$|\psi(0)\rangle=|E_1\rangle$
       with maximum pulse strength, $U_o=13.0$ for the ladder processes
in (a) and (b) and for the lambda process (c).
For the ladder process, the frequencies of both pulses are
$\omega_f=9\omega_o$ and $\omega_s=15\omega_o$ and (a) $t_{tot}=600$ and
(b) $t_{tot}=6000$.
For the lambda process, the frequencies of both pulses are
$\omega_f=7\omega_o$ and $\omega_s=15\omega_o$ and (c) $t_{tot}=6000$.
The numbers attached to each curve show the
components of the transition
probability in terms of the unperturbed energy levels.}
\label{fig.15}
\end{figure}

\begin{figure}
\caption{Husimi plot for the Floquet state $A$ for maximum
pulse   strength $U_o=13.0$ and frequencies
$\omega_f=9\omega_o$ and $\omega_s=15\omega_o$ at time
$t_{fix}=13/20 t_{tot}$, just before the avoided crossing at
$t_{fix}=\tau_{V}$.
This state covers the entire chaotic region of the underlying
classical phase space from $J=1$ to
$J=9$ and the highly distorted mixed region at $J=10$.}
\label{fig.16}
\end{figure}

\begin{figure}
\caption{Strobe plots of the action-angle variables $(J,\theta)$ for
the infinite square well system with
pulse  pulse amplitudes $U_o=13.0$ and frequencies
$\omega_f=7\omega_o$ and $\omega_s=15\omega_o$,
respectively.  Strobe plots are shown at times (a) $t_{fix}=t_f$, (b)
$t_{fix}=t_c$, (c) $t_{fix}=\tau_{IV}={0.62}t_{tot}$ and
(d) $t_{fix}=t_s$.  The first primary resonance  from
the first pulse is located at
$J=3.5$  and the first primary resonance from the second pulse is located
at $J=7.5$.}
\label{fig.17}
\end{figure}

\end{document}